\documentclass[intlimits,twoside,a4paper]{article}
\usepackage{amsmath,amssymb}
\usepackage{graphicx}
\usepackage[T2A]{fontenc}
\usepackage[cp1251]{inputenc}

\usepackage[eqsecnum]{cmpj2}

\newcommand{\be}{\begin{equation}}
\newcommand{\ee}{\end{equation}}
\newcommand{\bea}{\begin{eqnarray}}
\newcommand{\eea}{\end{eqnarray}}

\DeclareMathOperator{\erfc}{erfc}
\DeclareMathOperator{\erf}{erf}

\issue{2016}{19}{4}{43602}
\doinumber{10.5488/CMP.19.43602}

\title[Ring polymers in confined geometries]{Ring polymers in confined geometries}
\author[Z. Usatenko, J. Halun, P. Kuterba]{Z. Usatenko\refaddr{1}, J. Halun\refaddr{2}, P. Kuterba\refaddr{2}}
\addresses{\addr{1} Institute of Physics, Cracow University of Technology, 30-084 Cracow, Poland
\addr{2}Institute of Physics, Jagiellonian University, 30-348
Cracow, Poland}

\authorcopyright{Z. Usatenko, J. Halun, P. Kuterba, 2016}
\date{Received June 25, 2016, in final form September 6, 2016}

\begin{document}
\maketitle
\begin{abstract}
The investigation of a dilute solution of phantom ideal ring
polymers and ring polymers with excluded volume interactions (EVI)
in a good solvent confined in a slit geometry of two parallel
repulsive walls and in a solution of colloidal particles of big size
was performed. Taking into account the correspondence between the
field theoretical $\phi^4$ $O(n)$-vector model in the limit $n\to 0$
and the behaviour of long-flexible polymers in a good solvent, the
correspondent depletion forces and the forces which exert phantom
ideal ring polymers and ring polymers with EVI on the walls were
obtained in the framework of the massive field theory approach at
fixed space dimensions $d$=3 up to one-loop order. Besides, taking
into account the Derjaguin approximation, the depletion forces
between big colloidal particle and a wall and in the case of two big
colloidal particles were calculated. The obtained results indicate
that phantom ideal ring polymers and ring polymers with EVI due to
the complexity of chain topology and for entropical reasons
demonstrate a completely different behaviour in confined geometries
compared with linear polymers.
\keywords colloidal systems,critical
phenomena, polymers, phase transitions
\pacs{64.70.pv, 61.25.he,
67.30.ej, 68.35.Rh, 64.70.qd}
\end{abstract}
\renewcommand{\theequation}{\arabic{equation}}
\section*{}
\vspace{-3mm}
\looseness=-2 As it was shown in a series of the atomic force spectroscopy (AFM)
experiments \cite{Marek05,WitzRechendorffAdamcikDietler11},
biopolymers such as DNA very often present a ring topology. Such
a situation takes place, for example, in the case of Escherichia coli
(E.coli) bacteria with a chromosome which is not a linear polymer,
but has a ring topology \cite{BergTymoczkoStryer02}. The biopolymers
of DNA of some viruses such as bacteriophages $\lambda$ that infect
bacteria oscillate between linear and ring topology
\cite{ArsuagaVazquezTriguerosSumnersRoca02,MatthewsLouisYeomans09}.
The physical effects arising from confinement and chain topology
play a significant role in the shaping of individual chromosomes and
in the process of their segregation, especially in the case of
elongated bacterial cells \cite{JunMulder06}. The behaviour of
linear ideal and real polymers with excluded volume interaction
(EVI) in a good solvent confined in a slit of two parallel repulsive
\cite{SHKD01,RU09}, inert or mixed walls is well understood
\cite{RU09}. Unfortunately, the physics of confined ideal ring
polymers and ring polymers with EVI effects is still unclear. Ring
polymers with specified knot type were chemically synthesized a long
time ago \cite{Dietrich-BucheckerSauvage89}. Ring topology of
polymers influences the statistical mechanical properties of these
polymers, for example the scaling properties
\cite{Dobay03,Ercolini07} and shape
\cite{AlimFrey07,Rawdon08,WitzRechendorffAdamcikDietler11} because
it restrains the accessible phase space. An interesting point which was
confirmed by numerical studies in \cite{RensburgWhittington90} is
that longer ring polymers are usually knotted with higher frequency
and complexity. In \cite{Quake94}, it was established that ring
polymers with more complex knots are more compact and have a smaller
radius of gyration and this decreases their ability to spread out
under confinement. The results of Monte Carlo simulations performed
in \cite{JvanRensburg07} suggest that the knotted ring polymers will
exert higher entropic forces on the walls of the confining slit than
the unknotted or linear polymers. In \cite{JvanRensburg07} it was stated
that the knotted ring polymers expanded as the width of the slit
increased in contrast to the behaviour of unknotted (or linear)
polymers whose size showed a plateau after a certain width of slit
was reached. The entropic force exerted on the walls arising from
confinement to a slit of a knotted ring polymer was calculated using
a bead-spring model by Matthews et al. in
\cite{MatthewsLouisYeomans09}. It was found
\cite{MatthewsLouisYeomans09} that in the case of a narrow slit, more
complex knot types in a ring polymer exert higher forces on the
confining walls of the slit in comparison to unknotted polymers of
the same length, and for the relatively wide slits, the opposite
situation takes place. Confining ring polymer to a slab results in
a loss of configurational entropy and leads to the appearance of a
repulsive force which depends on the entanglements between two
walls of the confining slab, as it was shown in the framework of a
new numerical approach based on the generalized atmospheric sampling
(GAS) algorithm for lattice knots in
\cite{GasumovaRensburgRechnitzer12}. Thus, at the moment most of the
papers devoted to the investigation of the behaviour of ring polymers
compressed in confined geometries like slit or squeezed by a force
in a slab of two parallel walls deal with numerical
methods and present analytical results are incomplete. The above
mentioned arguments stimulate us to apply one of the powerful
analytical methods referred to as the massive field theory approach in a
fixed space dimensions $d<4$ (see \cite{Par80,Parisi}) for the
investigation of ring polymers confined to a slit geometry of two
parallel walls or immersed in a solution of big mesoscopic colloidal
particles of different size. This method, as it was shown in the
series of papers \cite{DSh98,RU09,U11}, provides a good agreement with
the experimental data and with the results of Monte Carlo
simulations.
 We consider a dilute polymer solution, where different
polymers do not overlap and the behaviour of such polymer solution
can be described by a single polymer. As it is known, taking into
account the polymer-magnet analogy developed by de Gennes
\cite{deGennes1,deGennes2}, the scaling properties of long-flexible polymers in
the limit of an infinite number of steps $N$ may be derived from a
formal $n \to 0$ limit of the field theoretical $\phi^4$ $O(n)$-vector model at its critical point. In this case, the $1/N$ value
plays the role of a critical parameter analogous to the reduced
critical temperature in magnetic systems. Besides, we assume that
the surfaces of the confining slit are impenetrable. It means that
the correspondent potential $U({\tilde{z}})$ of the interaction
between the monomers of a polymer chain and a wall tends to infinity
$U({\tilde{z}})\to\infty$ when the distance ${\tilde{z}}$ between a
wall and polymer is less than monomer size $l$. The deviation from
the adsorption threshold $[c\propto(T-T_{\text{a}})/T_{\text{a}}]$ (where $T_{\text{a}}$ is
adsorption temperature) changes the sign at the transition between the
adsorbed (the so-called normal transition, $c<0$) and the
nonadsorbed state (ordinary transition, $c>0$) \cite{D86,E93} and it
plays the role of a second critical parameter. The value $c$
corresponds to the adsorption energy divided by $k_{\text{B}}T$ (or the
surface enhancement in field theoretical treatment). The adsorption
threshold for long-flexible polymers takes place, where $1/N\to 0$
and $c\to 0$. As was mentioned by de Gennes \cite{deGennes1,deGennes2}, the
partition function $Z({\bf{\tilde{x}}},{\bf{\tilde{x}'}})$ of a
single polymer with two ends fixed at ${\bf{\tilde{x}}}$ and
${\bf{\tilde{x}'}}$ is connected with the two-point correlation
function
$G^{(2)}({\bf{\tilde{x}}},{\bf{\tilde{x}'}})=\langle{\vec{\phi}({\bf
\tilde{x}})}{\vec{\phi}}({\bf{\tilde{x}}}')\rangle$  in  $\phi^4$ $O(n)$-vector model for field $\vec{\phi}({\bf{\tilde{x}}})$ with the
components $\phi_i({\tilde{x}})$, $i=1,\ldots,n$ [and
${\bf{\tilde{x}}}=({\bf{\tilde{r}}},{\tilde{z}})$] via the
$\mu_{0}^{2}\to L_{0}$ inverse Laplace transform: $
Z({\bf{\tilde{x}}},{\bf{\tilde{x}}}';N,v_{0})={\cal
IL}_{\mu_{0}^2\to L_{0}}(\langle{\vec
\phi}({\bf{\tilde{x}}}){\vec\phi}({\bf{\tilde{x}'}})\rangle|_{n\to 0})$ in
the limit, where the number $n$ of components tends to zero. The
conjugate Laplace variable $L_{0}$ has the dimension of length
squared and is proportional to the total number of monomers $N$
which form the polymer. The effective Ginzburg-Landau-Wilson
Hamiltonian describing the system in semi-infinite ($i=1$) or
confined geometry of two parallel walls ($i=1,2$) is  \cite{D86}:
\be {\cal H}[{\vec \phi,\mu_{0}}] =\int \rd^{d}{\tilde{x}}\left[
\frac{1}{2} \left( \nabla{\vec{\phi}} \right)^{2}
+\frac{{\mu_{0}}^{2}}{2} {\vec{\phi}}^{2} +\frac{v_{0}}{4!}
\left({\vec{\phi}}^2 \right)^{2} \right]
+\sum_{i=1}^{2}\frac{c_{i_{0}}}{2} \int \rd^{d-1}{\tilde{r}}
{\vec{\phi}}^{2}, \label{hamiltonianslit}\ee where the conjugate
chemical potential $\mu_0$ is the ``bare mass'' in field-theoretical
treatment, $v_0$ is the ``bare coupling constant'' which characterizes
the strength of
 EVI. In the case of slit geometry, the walls are located at the
distance $L$ one from another in ${\tilde{z}}$-direction in such
way that the surface of the bottom wall is located at
${\tilde{z}}=0$ and the surface of the upper wall is located at
${\tilde{z}}=L$. Each of the two surfaces is characterized by a
certain surface enhancement $c_{i_{0}}$, where $i=1,2$. In the case
when the ends of polymer ${\bf{\tilde{x}}}$ and ${\bf{\tilde{x}}'}$
in partition function
$Z({\bf{\tilde{x}}},{\bf{\tilde{x}}'};N,v_{0})$ coincide, such
partition function corresponds to the partition function of a
phantom ring polymer, i.e., a ring polymer where we perform the
summation over all possible knot structures. The fundamental
two-point correlation function of the Gaussian theory corresponding
to the effective Ginzburg-Landau-Wilson Hamiltonian
(\ref{hamiltonianslit}) in a mixed ${\bf{\tilde{p}}},{\tilde{z}}$
representation is: $ G^{(2)}_{ij}({\bf{\tilde{p}}},{\bf{\tilde{
p}'}};{\tilde{z}},{\tilde{z}'})= (2\pi)^{d-1}\delta_{ij}\delta ({\bf
\tilde{p}}+{\bf{\tilde{p}}'})\, {\tilde G}_{\parallel}({\bf
{\tilde{p}}};{\tilde{z}},{\tilde{z}'};\mu_{0},c_{1_{0}},c_{2_{0}},L)$,
where the free propagator $ {\tilde G}_{\parallel}({\bf
{\tilde{p}}};{\tilde{z}},{\tilde{z}'};\mu_{0},c_{1_{0}},c_{2_{0}},L)$
of the model (\ref{hamiltonianslit}) in the case of ring polymer
with coinciding ends ${\bf{\tilde{x}}}={\bf{\tilde{x}}'}={\bf{x}}$
can be obtained by analogy as it was done in \cite{RU09} for linear
polymer. It should be taken into account that we have ring polymer
of length $L_{0}$ in a slit geometry of two parallel walls with
fixed position of one monomer at point ${\bf x}=({\bf r},z)$ by
analogy as it was proposed in \cite{E93} for description of ring
polymer adsorption. Such representation assumes that the respective
correlation function for ring polymer depends on ${\bf x}=({\bf
r},z)$, e.g., on the position of some monomer which can be the first
and the last monomer in the ring.
 Here, ${\bf{\tilde{p}}},{\bf{\tilde{p}}'}$ are the values of parallel momentum associated with
$d-1$ translationally invariant directions in the system. The
interaction between the polymer and the walls is implemented by the
Dirichlet-Dirichlet boundary conditions (D-D b.c.) (see \cite{D86,SHKD01,RU09}): $c_{1}\to +{\infty}\,$, $c_{2}\to
+{\infty}$ or ${\vec \phi}({\bf {r}},0)={\vec \phi}({\bf
{r}},L)=0$. We consider the dilute solution of phantom ideal ring
polymers and ring polymers with EVI immersed in a slit geometry of
two parallel repulsive walls and permit the exchange of polymer coils
between the slit and the reservoir. The polymer solution in the slit
is in equilibrium contact with an equivalent solution in the
reservoir outside of the slit. We follow the thermodynamic
description of the problem as it was given in \cite{SHKD01,RU09} and
perform the calculation in the framework of the grand canonical
ensemble where the chemical potential~$\mu$ is fixed. As it was
shown in \cite{SHKD01}, the free energy of interaction between
the walls in such a grand canonical ensemble is defined as the
difference of the free energy of an ensemble where the separation of
the walls is fixed at a finite distance $L$ and where the walls are
separated infinitely far from each other: $ \delta F^{\text R} = -k_{\text{B}} T
\,{\cal N}\,\ln\big\{{{\cal Z}_{\parallel}^{\text R} (L)}/[{{\cal
Z}_{\parallel}^{\text R}(L\to\infty)}]\big\}$,
 where $\cal{N}$ is the total number of polymer coils in the solution and $T$ is
the temperature. The ${\cal Z}_{\parallel}^{\text R}(L)$ value is the
partition function of one ring polymer located in a volume $V$
containing two walls at a distance $L$. The correspondent {\it
reduced free energy of interaction} $\delta f$ per unit area $A=1$
for the case of ring polymer confined in a slit geometry of two
parallel walls after performing Fourier transform in the direction
parallel to the surfaces and integration over $\rd^{d-1}r$ (which
gives us the delta function $\delta ({\bf p})$ and leads to ${\bf
p}=0$ in the respective correlation functions after integration over
$\rd^{d-1} p$) may be written in the form (see \cite{RU09}):
\be
\delta f^{\text R} = \frac{\delta{F^{\text R}}}{n_{\text p}k_{\text{B}}T}=L - \int_{0}^{L} \rd
z \frac{{\hat{\cal Z}}_{\text I}^{\text R} (z)}{{\hat{\cal Z}}_{\text b}} +
\int_{0}^{\infty}\rd z \left[\frac{{\hat{\cal
Z}}_{\text{HS}_{1}}^{\text R}(z)}{{\hat{\cal Z}}_{\text b}}-1\right]+\int_{0}^{\infty}\rd z
\left[\frac{{\hat{\cal Z}}_{\text{HS}_{2}}^{\text R}(z)}{{\hat{\cal
Z}}_{\text b}}-1\right].
\label{dfringz}
\ee
Here, $n_{\text p}={\cal N}/V$ is the
polymer density in the bulk solution, and ${\hat{\cal
Z}}_{\text b}={\cal{IL}}_{\mu_{0}^{2}\rightarrow R_{x}^2/2}\big(\frac{1}{2\mu_{0}}\big)$.
The functions ${\hat{\cal Z}}_{\text I}^{\text R}(z)$ and ${\hat{\cal
Z}}_{\text{HS}_i}^{\text R}(z)$ are equal to:
\[{\hat{\cal Z}}_{\text I}^{\text R}(z)={\cal
IL}_{\mu_{0}^2\to L_{0}} G^{(2)}({\bf p}=0;z,z)\big|_{n\to 0}\,, \qquad \text{and} \qquad
{\hat{\cal Z}}_{\text{HS}_{i}}^{\text R}(z)={\cal IL}_{\mu_{0}^2\to L_{0}}
G^{(2)}_{\text{HS}_{i}}({\bf p}=0;z,z)\big|_{n\to 0}\,,\]
where $G^{(2)}({\bf
p}=0;z,z)$ and $G^{(2)}_{\text{HS}_{i}}({\bf p}=0;z,z)$ with $i=1,2$ are
correlation functions of the model equation~(\ref{hamiltonianslit})
describing the system in a slit geometry of two parallel walls and
in semi-infinite geometry, respectively (see \cite{DSh98,RU09}).
Dividing  the {\it reduced free energy of interaction} $\delta
f^{\text R}$ by another relevant length scale, for example, the size of
the polymer in bulk, e.g. $R_x$ (where
$R_{\text g}^{2}\,=\chi^2_d R_x^2/{2}$, and $\chi_d$ is a universal
numerical prefactor depending on the dimension $d$ of the system
(see \cite{E93},\cite{CJ90}) and $R_{\text g}$ is the radius of
gyration), yields the universal scaling function for the {\it
depletion interaction potential} $\Theta^{\text R}(y) = \delta
f^{\text R}/{R_x}$, where $ y = L/R_{x} $. The resulting scaling function
for the {\it depletion force} between the two walls induced by the
polymer solution is denoted as (see \cite{RU09}):
\be \Gamma^{\text R}(y)
=-\frac{\rd (\delta
f^{\text R})}{\rd L}=-\frac{\rd\Theta^{\text R}(y)}{\rd y}\,.
\label{Gamma}
\ee
As it is
known \cite{E93}, the resulting force exerted on the surfaces of a
confining slit by polymer is equal to the resulting {\it depletion
force} with the opposite sign: $K^{\text R}=\rd (\delta f^{\text R})/\rd L$.

\begin{figure}[!t]
\vspace{-3mm}
\begin{center}
\includegraphics[width=7.0cm]{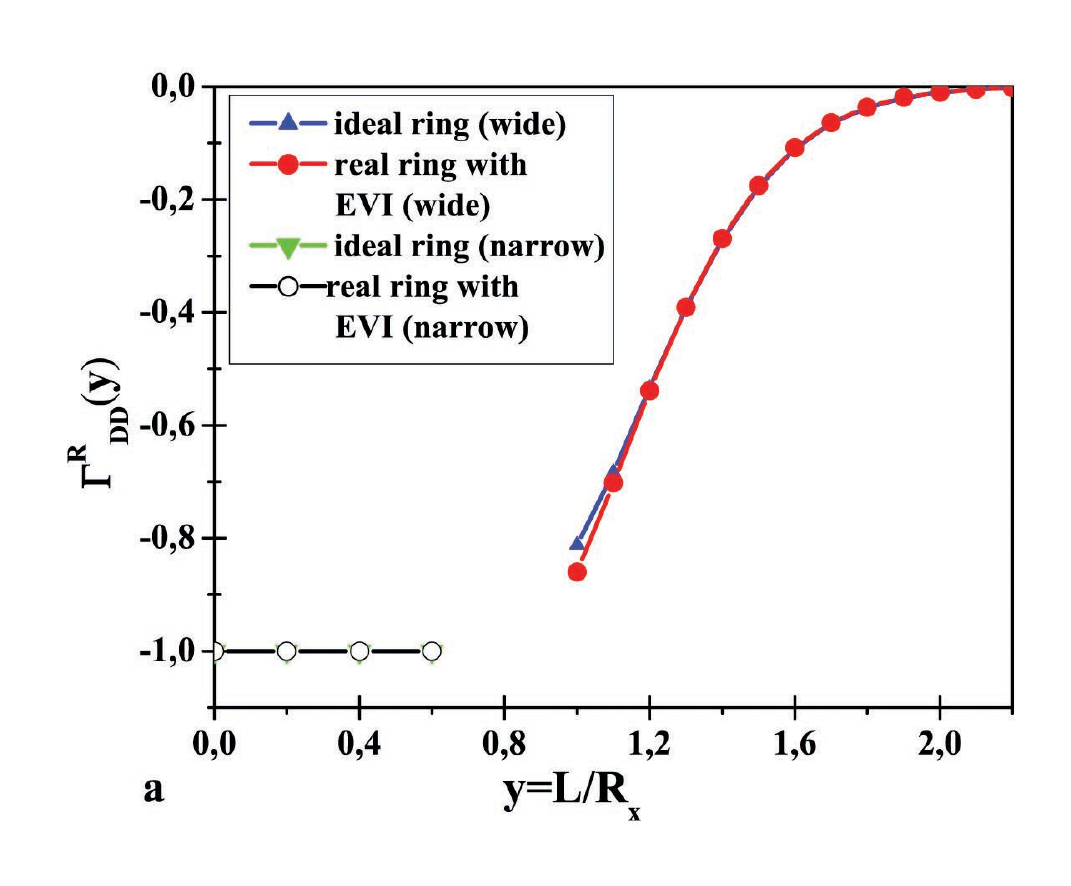}
\includegraphics[width=7.0cm]{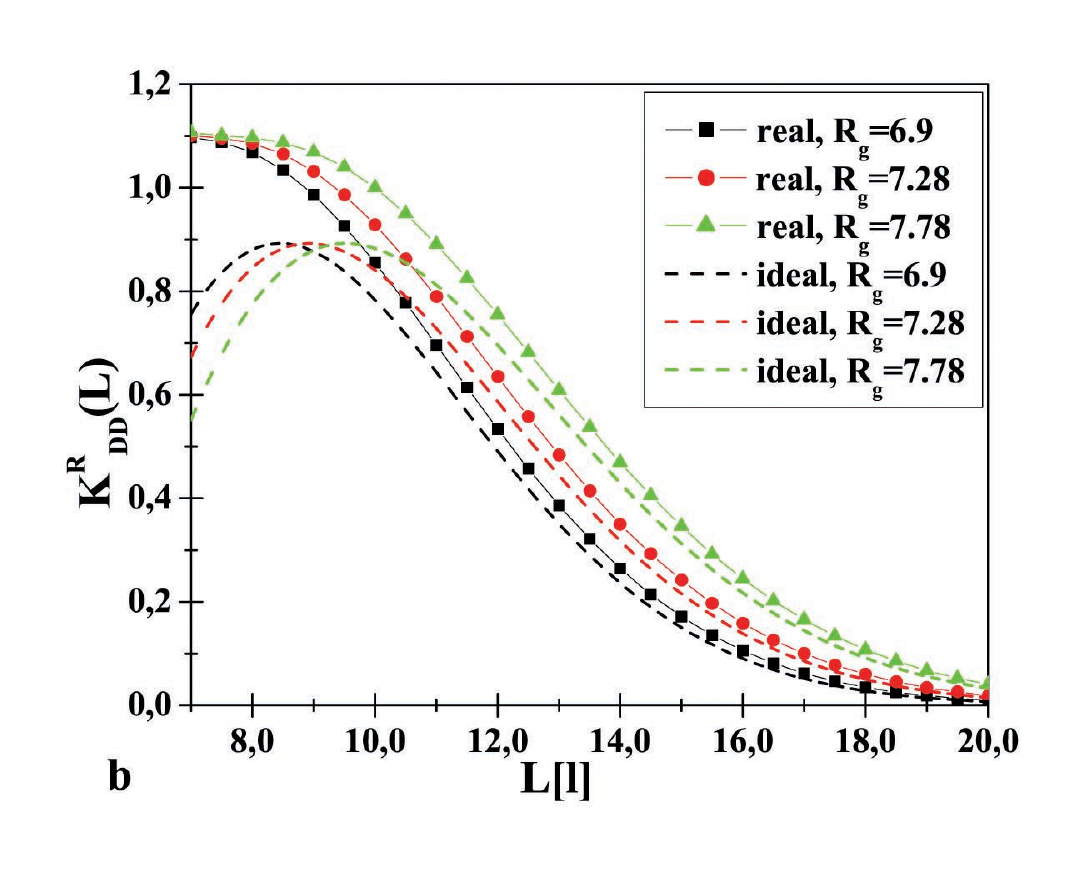}
\vspace{-5mm}
\caption{(Color online) (a) The scaling functions $\Gamma^{\text R}_{\text{DD}}(y)$ for phantom
ideal ring and ring polymer with EVI in a good solvent immersed
between two repulsive walls; (b) The functions $K^{\text R}(L)$ for
phantom ideal ring polymer and ring polymer with EVI in a good
solvent immersed between two repulsive walls for different values of
the radius of gyration: $R_{\text g}(12_{1})=6.9\pm 0.01 [l]$ (black lines
and black lines with squares); $R_{\text g}(9_{1})=7.28\pm 0.01 [l]$ (red
dashed lines and red lines with dots); $R_{\text g}(6_{1})=7.78\pm 0.01
[l]$ (green dashed lines and green lines with triangles). }
\label{fig:gamma_DD}
\end{center}
\vspace{-2mm}
\end{figure}

\begin{figure}[!h]
\vspace{-3mm}
\begin{center}
\includegraphics[width=7cm]{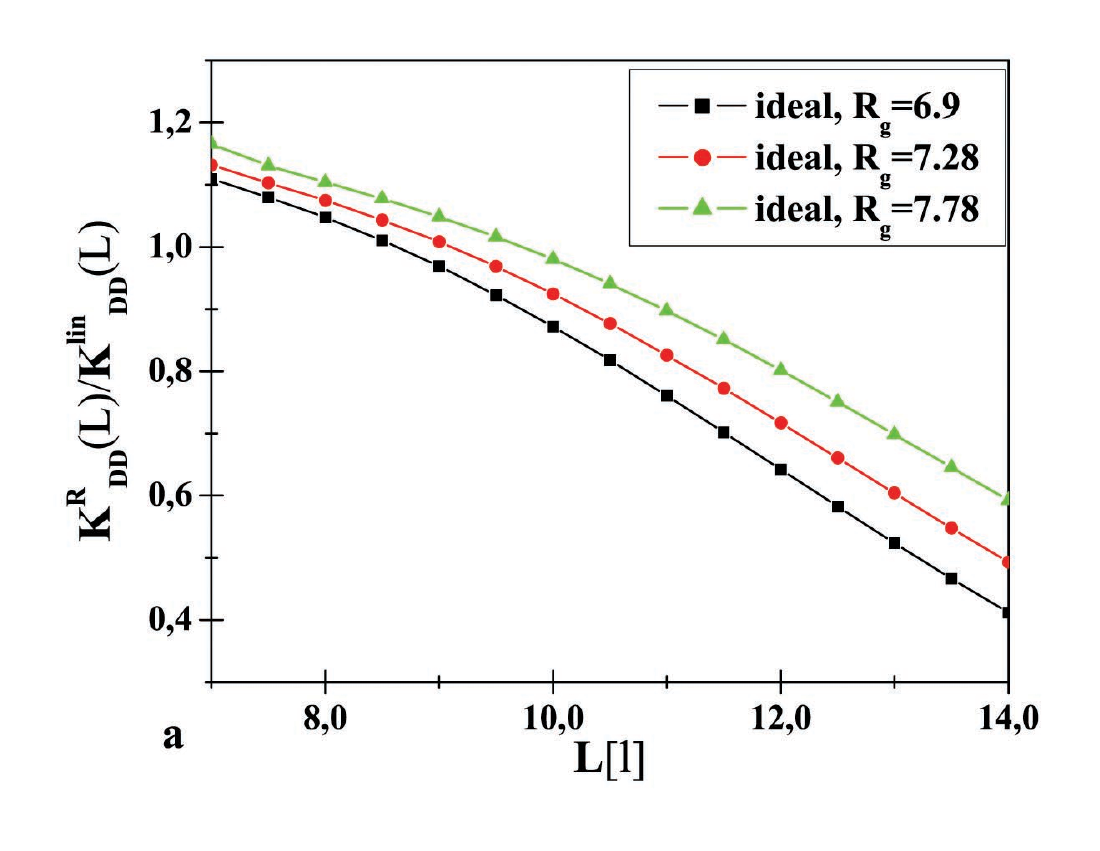}
\includegraphics[width=7.3cm]{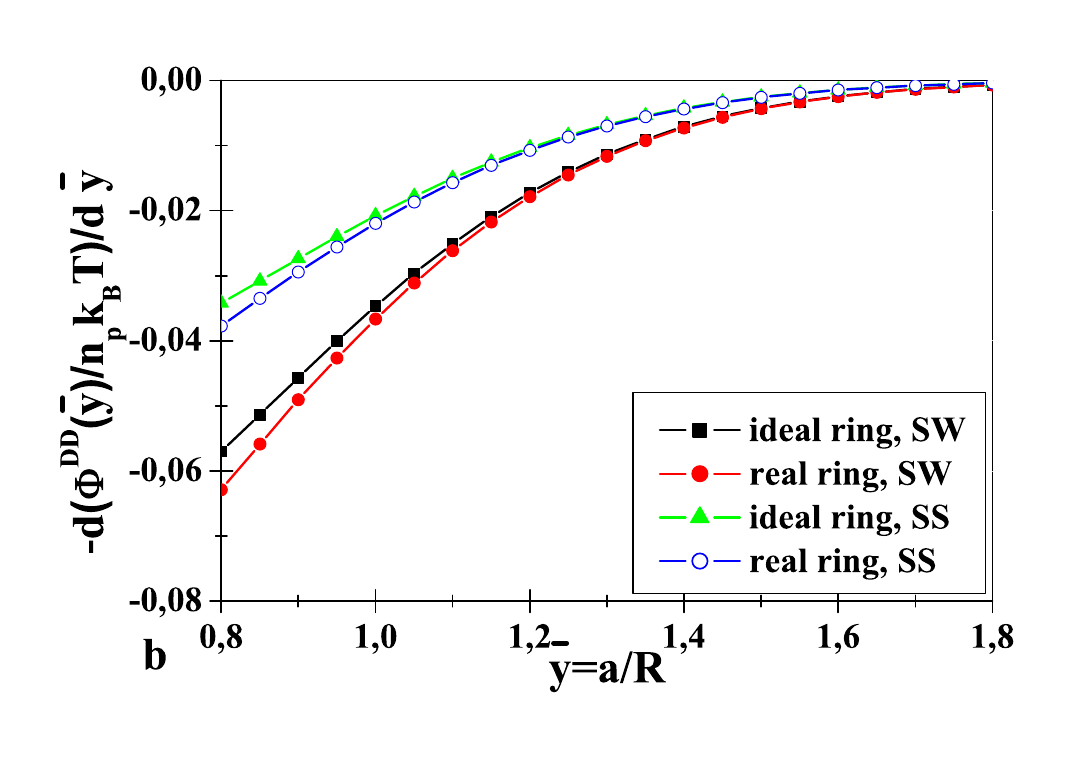}
\vspace{-5mm}
\caption{(Color online) (a) The ratio $K^{\text R}(L)/K^{\text{lin}}(L)$ as function of the
distance $L$ (in $l$ units) between the walls for different values
of $R_{\text g}$; (b) The functions
$-\rd[\Phi^{\text R}_{\text{DD}}(\tilde{y})/(n_{\text p}k_{\text{B}}T)]/\rd{\tilde{y}}$ for phantom
ideal ring polymer and ring polymer with EVI in a good solvent
immersed between colloidal particle and wall as well as between two
colloidal particles.} \label{fig:gamma_SPSS}
\end{center}
\vspace{-2mm}
\end{figure}

Let us consider at the beginning the case of phantom ideal ring
polymer under $\Theta$-solvent condition trapped in the slit
geometry of two parallel repulsive walls. Taking into account that
in the case of wide slit region we have $y\gtrsim 1$ for the scaling
function of the {\it depletion force} $\Gamma^{\text R}$ we obtain:\be
\Gamma^{\text{R,id}}_{\text{DD}}(y)=2\re^{-2y^2}-8y^{2}\re^{-2y^2}.\label{ThGamDD} \ee
The asymptotic solution for the {\it depletion force} in the narrow
slit region, where $y\ll1$, simply becomes:
$\Gamma^{\text{R,id}}_{\text{DD,\,narr}}{\approx}-1$. The obtained results for the
{\it depletion force} $\Gamma^{\text{R,id}}_{\text{DD}}$ are presented in figure~\ref{fig:gamma_DD}~(a)
by blue lines with triangles for a wide slit region and by green
lines with triangles for a narrow slit region, respectively.
It should be mentioned that the quantity ${\Gamma}$ is normalized
to the overall polymer density $n_{\text p}$. Thus, the above result simply
indicates that the force is entirely induced by free chains
surrounding the slit, or, in other words, by the full bulk osmotic
pressure from the outside of the slit. We can state that in the case
of very narrow slit with two repulsive walls, the ring polymers would
pay a very high entropy to stay in the slit or even enter it.
 As the next step, let us consider the dilute solution of ring polymers
with EVI in a good solvent immersed in a slit geometry of two
parallel repulsive walls. As it is known \cite{deGennes1,deGennes2}, in a good
solvent the effects of the EVI between monomers play a crucial role
so that the polymer coils occupy a large space compared to the case of
ideal polymers. The calculations of the correspondent partition
functions ${\hat{\cal Z}}_{\text I}^{\text R}(z)$ and ${\hat{\cal
Z}}_{\text{HS}_{i}}^{\text R}(z)$, which permit to obtain the reduced free energy
of interaction $\delta f^{\text R}$ in equation~(\ref{dfringz}), are connected
with the calculations of the correspondent correlation functions
$G^{(2)}({\bf p}=0;z,z)$ and $G^{(2)}_{\text{HS}_{i}}({\bf p}=0;z,z)$ with
$i=1,2$ where the terms describing the EVI effects are taken into
account via using perturbation treatment in the framework of the
massive field theory approach in a fixed space dimensions $d=3$ up
to one-loop order approximation. In order to make the theory UV
finite in RG sense directly in $d=3$ space dimensions, we perform a
standard mass renormalization $\mu^{2}_{0}=\mu^{2}-\delta
\mu^{2}_{0}$ and the coupling constant renormalization $v_{0}=\mu v$
of the above mentioned correlation functions by analogy as it was
proposed by Parisi \cite{Parisi}. Besides, the surface enhancement
renormalization $c_{i_{0}}=c_{i}+\delta c_{i}$ of the correspondent
correlation functions in the case of D-D b.c. reduces to an additive
renormalization as it took place in the case of semi-infinite
geometry \cite{DSh98} and slit geometry (see \cite{RU09}). The
correspondent expression for the scaling function of the {\it
depletion force} between two repulsive walls in the case of wide
slit region $y\gtrsim 1$ is:
\be
{\Gamma}^{\text{R,\,real}}_{\text{DD}}(y)=2\re^{-2y^2}- 8y^{2}\re^{-2y^2} -\frac{\tilde
v}{2}\re^{-2y^2}\left[\frac{1}{y^2}+2\sqrt{\pi}\psi^{(0)}\left(-\frac{1}{2}\right)-
D-4yB(y)\right],
\label{GammarealDD}
\ee
 where
$D=\frac{1}{4}\big[10-2\ln
2-2\psi^{(0)}(-\frac{1}{2})+\sqrt{\pi}\psi^{(0)}(\frac{1}{2})-
2\gamma_{\text E}\big]$,
$B(y)=\big[2y\sqrt{\pi}\psi^{(0)}(-\frac{1}{2})-\frac{1}{y}\big]$ and
$\gamma_{\text E}=0.577$ is the Euler's constant, $\psi^{(0)}(z)$  is
the digamma function, $\erfc(z)=1-\erf(z)$ is the complementary error
function. Besides, the $v=b_{n}{\tilde v}$ value was introduced with
$b_{n}=\frac{6}{n+8}\frac{(4\pi)^{3/2}}{\Gamma(1/2)}$ and the
calculations are performed at the correspondent fixed point ${\tilde
v}^{*}=1$ in the limit $n\to 0$.  In the region of a very narrow slit
$y\ll1$ the {\it depletion force} is:
$\Gamma^{{\text{R,\,real}}}_{\text{DD,\,narr}}{\approx}-1$. The results of calculations
for $\Gamma^{{\text{R,\,real}}}_{\text{DD}}$ are presented in figure~\ref{fig:gamma_DD}~(a) by the red
line with circles in a wide slit region and by the black line with
open circles in a narrow slit region, respectively. As it is easy to
see from figure~\ref{fig:gamma_DD}~(a), the obtained results for the scaling function
for the {\it depletion force} in the case of ring polymer immersed
between two repulsive walls are characterized by a completely
different behaviour compared to the case of linear polymer (see
\cite{RU09}). The phantom ring polymer, due to the complexity of
chain topology and for entropical reasons, tends to escape from the
space between two repulsive walls and it leads to the attractive
{\it depletion force} between the confining walls. Besides, we
observe [see figure~\ref{fig:gamma_DD}~(a)] that the absolute value of the scaling
function for the {\it depletion force} for phantom ring polymers in
the wide slit region is smaller than for linear polymer chains (see
\cite{RU09}) and decreases as the width of the slit increases. In
figure~\ref{fig:gamma_DD}~(b) we present the results for the entropically induced
force $K^{\text R}$ (as functions of $L$ for different values of radius of
gyration $R_{\text g}$) which exerts phantom ideal ring polymer and ring
polymer with EVI on the confining two repulsive walls. It should be
mentioned that calculations, presented in figure~\ref{fig:gamma_DD}~(b) were performed
for different values of radius of gyration
\cite{MatthewsLouisYeomans09}: $R_{\text g}(12_{1})=6.9\pm 0.01 [l]$,
$R_{\text g}(9_{1})=7.28\pm 0.01 [l]$, and $R_{\text g}(6_{1})=7.78\pm 0.01 [l]$
which correspond to the ring polymers with different knot types:
$12_{1};9_{1};6_{1}$. Here, $C_{p}$ is a standard notation
\cite{OrlandiniWhittington07}, where $C$ denotes the minimum number
of crossings in any projection on a plane and $p$ is used in order
to distinguish the knot types with the same $C$. It should be mentioned
that the region of validity of the obtained results for the entropically
induced forces $K^{\text R}$ for phantom ideal ring polymers [see figure~\ref{fig:gamma_DD}~(b)]
is defined by the value of $y=L/R_{x}\gtrsim 1$ with
$R_{x}=\sqrt{2}R_{\text g}$. We observed that in the wide slit region, ring
polymers with less complex knot types (with bigger radius of
gyration) in a ring topology exert higher forces on the confining
walls [see figure~\ref{fig:gamma_DD}~(b)] at the same length~$L$. In figure~\ref{fig:gamma_SPSS}~(a), we
present the ratio $K^{\text R}_{\text{DD}}/K^{\text{lin}}_{\text{DD}}$ of force for ring
polymer chains and corresponding force for linear polymer chains in
a slit geometry of two parallel repulsive walls as function of the
distance~$L$ between the walls for different values of the radius of
gyration $R_{\text g}$. Our results for entropically induced force $K^{\text R}$
are in agreement with the previous results obtained by Matthews et~al.
in \cite{MatthewsLouisYeomans09} using a bead-spring model. The
difference between our results and the results obtained by Matthews
et~al. arises due to the fact that in \cite{MatthewsLouisYeomans09}
calculations were performed for relatively short polymer chains with
polymer length of order $N\sim 300$ units. Taking into account the
Derjaguin approximation \cite{D34}, which describes the sphere of
the big colloidal particle by a superposition of immersed plates
with local distance from the wall (or from other particle), we
performed calculations of the {\it depletion force}
$-\rd[{\Phi^{\text R}_{\text{DD}}}({\tilde{y}})/(n_{\text p}k_{\text{B}}T)]/\rd{\tilde{y}}$
between colloidal particle and a wall (or between two colloidal
particles) in a dilute solution of ring polymers. It should be
mentioned that ${\Phi^{\text R}_{\text{DD}}}({\tilde{y}})/(n_{\text p}k_{\text{B}}T)$ is equal
to $2\pi{\tilde R}
R^{2}_{x}\int_{\tilde{y}}^{\infty}\rd y\Theta^{\text R}(y)$ with
${\tilde{y}}=a/\sqrt{2}R_{\text g}$ (where $a$ is the minimal distance
between the sphere and the wall) and ${\tilde R}=R$ for the case of
big colloidal particle of radius $R\gg L$ (and $R\gg R_{\text g}$) near the
wall and ${\tilde R}=R_{1}R_{2}/(R_{1}+R_{2})$ for the case of two
colloidal particles with different radius $R_{1}\neq R_{2}$ (when
$R_{i}\gg L$ and $R_{i}\gg R_{\text g}$, $i=1,2$).
 The results of calculations are presented in figure~\ref{fig:gamma_SPSS}~(b) and indicate that the absolute
value of the {\it depletion force} which arises between two
colloidal particles is smaller than between particle and wall.
Besides, if the number of coils $\cal{N}$ can be fixed, then the total
free energy  of the polymer solution within the slit in the
canonical ensemble can be obtained from our results for the grand
canonical free energy via the Legendre transform by analogy as it
took place in \cite{SHKD01,RU09} for linear polymers. The further
investigation of solution of ring polymers with EVI immersed in a
slit geometry of two inert walls or mixed walls as well as the
investigation of polymer-colloid interactions for different sorts of
colloids is the task of great interest which is under consideration.

\ukrainianpart

\title{Кільцеві  полiмери в обмежених середовищах}
\author{З.~Усатенко\refaddr{1}, Й.~Халюн\refaddr{2}, П.~Кутерба\refaddr{2}}
\addresses{
\addr{1} Інститут Фізики, Краківський Політехнічний Університет, 30-084 Краків, Польща
\addr{2} Інститут Фізики, Ягеллонський Університет, 30-348 Краків, Польща
}

\makeukrtitle

\begin{abstract}
\tolerance=3000%
Проведено дослідження розведених розчинiв iдеальних кільцевих та кільцевих полiмерiв з ефектами виключеного об’єму в доброму розчиннику, обмежених в щiлинi двох паралельних вiдштовхуючих поверхонь, а також у розчинi колоїдних частинок великого розмiру. Беручи до уваги вiдповiднiсть мiж теоретикопольовою  $\phi^4$~$O(n)$-векторною моделлю в границi $n\to 0$ i поведiнкою довгих гнучких полiмерiв в доброму розчиннику,  було отримано з використанням масивної теорiї поля при фiксованiй вимiрностi простору $d=3$ в рамках однопетлевого наближення вiдповiднi сили збіднення i сили, які  дiють між iдеальними  кільцевими та кільцевими полiмерами з ефектами виключеного об’єму і  стiнками щiлини. Окрiм того, беручи до уваги наближення Дєрягiна, розраховано сили збіднення мiж колоїдною частинкою i стiнкою, а також мiж двома колоїдними частинками великого розмiру. Отриманi результати вказують на те, що кільцевi полiмери через складність топологiї i з ентропiйних причин демонструють цiлком iншу поведiнку в обмежених середовищах
нiж лiнiйнi полiмери.

\keywords колоїдні системи, критичні явища, полімери, фазові
переходи
\end{abstract}
\end{document}